\documentclass[conference]{IEEEtran}
\IEEEoverridecommandlockouts

\usepackage{cite}
\usepackage{amsmath,amssymb,amsfonts}
\usepackage{algorithmic}
\usepackage{graphicx}
\usepackage{textcomp}
\usepackage{xcolor}
\def\BibTeX{{\rm B\kern-.05em{\sc i\kern-.025em b}\kern-.08em
    T\kern-.1667em\lower.7ex\hbox{E}\kern-.125emX}}
\begin{document}

\newcommand{\vv}{\vspace{2mm}}
\newcommand{\vvv}{\vspace{3mm}}
\newcommand{\vvvv}{\vspace{4mm}}
\newcommand{\hh}{\hspace{2mm}}
\newcommand{\hhh}{\hspace{3mm}}
\newcommand{\hhhh}{\hspace{4mm}}

\title{Machine Learning for Network Attacks Classification and Statistical Evaluation of Adversarial Learning Methodologies for Synthetic Data Generation\\
}

\author{\IEEEauthorblockN{Iakovos-Christos Zarkadis}
\IEEEauthorblockA{
\textit{University of Piraeus}\\
Athens, Greece \\
iakovos.zarkadis@gmail.com}
\and
\IEEEauthorblockN{Christos Douligeris}
\IEEEauthorblockA{\textit{Dept. of Informatics} \\
\textit{University of Piraeus}\\
Piraeus, Greece \\
cdoulig@unipi.gr}
}

\maketitle
\begin{abstract}
Supervised detection of network attacks has always been a critical part of network intrusion detection systems (NIDS). Nowadays, in a pivotal time for artificial intelligence (AI), with even more sophisticated attacks that utilize advanced techniques, such as generative artificial intelligence (GenAI) and reinforcement learning, it has become a vital component if we wish to protect our personal data, which are scattered across the web. In this paper, we address two tasks, in the first unified multi-modal NIDS dataset, which incorporates flow-level data, packet payload information and temporal contextual features, from the reprocessed CIC-IDS-2017, CIC-IoT-2023, UNSW-NB15 and CIC-DDoS-2019, with the same feature space. In the first task we use machine learning (ML) algorithms, with stratified cross validation, in order to prevent network attacks, with stability and reliability. In the second task we use adversarial learning algorithms to generate synthetic data, compare them with the real ones and evaluate their fidelity, utility and privacy using the SDV framework, $f$-divergences, distinguishability and non-parametric statistical tests. The findings provide stable ML models for intrusion detection and generative models with high fidelity and utility, by combining the Synthetic Data Vault framework, the TRTS and TSTR tests, with non-parametric statistical tests and $f$-divergence measures.
\end{abstract}


\section{Introduction}
In the era of AI, detecting the type of the attack becomes even more difficult. Machine learning algorithms have been implemented in many research papers and application areas \cite{b10} showing efficiency in tackling attacks across a variety of well-known intrusion detection system (IDS) datasets \cite{b38}, \cite{b24}, \cite{b29}, \cite{b30}, \cite{b32}. Nevertheless, it is necessary to not simply focus on achieving high results in classification metrics, but offering reliability and stability to an intrusion detection system. 
\par In the last few years adversarial attacks have gained popularity due to their sophisticated attacks, that utilize adversarial machine learning algorithms to inject perturbations in the data, in order to make a strong classifier unable to classify certain attacks correctly, thus misleading the IDS. 
\par Many types of algorithms, from linear models and SVM to neural networks and reinforcement learning have been proposed in \cite{b5}. Commonly used algorithms in IoT tabular data are logistic regression as a baseline for multi-class classification as well as more advanced models like deep neural networks and ensemble models \cite{b9}. More detailed studies have examined cross-dataset validation with LDA, Xgboost and decision trees \cite{b24}. Different frameworks have been widely used for data generation purposes like Synthcity \cite{b40} and SDV \cite{b26}, \cite{b8} offering high quality generators and evaluating metrics. The attention of researchers has recently been directed in many conditional architectures for synthetic network data generation for GANs and VAEs \cite{b20}, \cite{b21}, \cite{b22}, \cite{b25}, \cite{b27}, \cite{b31}, as well as in the emerging $f$-GANs and diffusion models \cite{b6}, \cite{b12}, \cite{b39} with the LLMs being top-tier models \cite{b37}. Furthermore, the specific non-parametric statistical tests, for means, covariance matrices and multivariate distribution comparison, have been used in some papers considering network data \cite{b14}, \cite{b17}, \cite{b19}, but may have never been combined to offer a complete view. Our focus, in this paper is not how to perform such attacks but on how the generation process evolves and test different adversarial algorithms, diffusion forests and LLM to create synthetic data \cite{b27}, \cite{b31} and to evaluate which are the most efficient ones, in terms of fidelity, utility and privacy. In this paper we use TRTS, TSTR tests, supervised distinguishability tests, $f$-divergences, privacy measurement \cite{b11} and combine for the first time the SDV framework \cite{b34}, with robust non-parametric statistical tests for comparing multivariate means, covariance matrices and the feature's joint multivariate distributions \cite{b14}, \cite{b17}, \cite{b19}.
\par This work is organized in two sections. In the first we present the classification of the attacks, where we discuss data collection, pre-processing steps and the ML models used for intrusion detection. In the second section we show the generation process of synthetic data, including a short review of the mathematical literature, the training of adversarial, diffusion and LLM algorithms and finally the evaluation procedure. 

\section{Machine Learning-Based NIDS Methodology}

\subsection{Data Collection}
For all the results that will be presented in the next sections we used the under-sampled version of the UMNIDS dataset, a newly created dataset, which was created by processing raw pcap files from CIC-IDS 2017, CIC-IOT 2023, UNSW-NB15 and CIC-DDoS 2017 \cite{b34}. UMNIDS includes flow-level and payload-based data with features that capture historical and temporal patterns in network traffic. This dataset includes around $170,000$ observations, $88$ features and $1$ target variable, which implies if a record is normal or if it is an attack in the network along with the type of the attack. Some that are included are Worms, Bots, DDoS, Infiltration and SQL-Injection. The target variable includes a total of $44$ unique values, $43$ of which specify the kind of attack and the one remaining the non-malicious records. One of the main issues with this dataset is the lack of records for some attacks, like Heartbleed and Infiltrations.

\subsection{Data Pre-Processing}
The pre-processing procedure started by omitting the duplicate values, then encoding the label variable, testing different methods like binning for certain numeric variables with a small number of unique values. We performed an in-depth statistical analysis that included the discovering of linear and non-linear relationships, hypothesis testing for existence of normality through QQ-Plots and goodness of fit tests like Kolmogorov-Smirnov and Anderson Darling, as well as regression analysis for some numeric features. We also examined the existence of outliers with IQR, Z-score, DBSCAN and Local-Outlier Factor and due the great number of outliers we preferred to keep them and use as a scaling method the Robust scaler, after we performed an $80/20$ train/test stratified split, with respect to the attack type variable. 
\par Furthermore a great deal of feature engineering techniques were used to increase the efficiency of the ML algorithms, separately for the training and testing data. To tackle the problem of the unbalanced dataset, we investigated different over-sampling and under-sampling techniques only for the training set to avoid data leakage, like synthetic minority oversampling (SMOTE), edited nearest neighbor (ENN) and adaptive synthetic sampling (ADASYN) that lead us to select the over-sampling approach to reduce the impact of information loss. Finally for the feature selection procedure we tested multiple selection techniques like recursive feature elimination, stepwise (forward-backward) elimination and boruta.

\subsection{Attacks Classification}
For the classification of the tabular data, we selected baseline and top-tier models from different families. The two baseline models that we used for the multi-class classification task are Logistic Regression and the stochastic gradient descent-based SVM. The tree-based model that we utilized is a Decision Tree. The ensemble models are LightGBM, XGBoost, XGBoost with Random Forest, Random Forest, Extra Trees, Gradient Boosting, and Histogram-Based Gradient Boosting. Finally we also chose to use the AutoGluon, an ensemble of neural networks, extra trees, XGBoost, LightGBM and linear models. Several hyperparameter optimization techniques were tested, such as grid-search, halving-grid-search, and Bayesian optimization, along with stratified $10$-fold cross validation. After the selection of the hyperparameters, each of the pre-mentioned models has been trained with stratified $10$-fold cross validation to observe their behavior without probability calibration. After that, we performed stratified $10$-fold cross validation with probability calibration, while holding a $4\%$ value range threshold for considering a model stable per metric, and obtained the results that follow for the training and testing set. 
\par
\begin{figure}[htbp]
\centerline{\includegraphics[width=0.95\columnwidth]{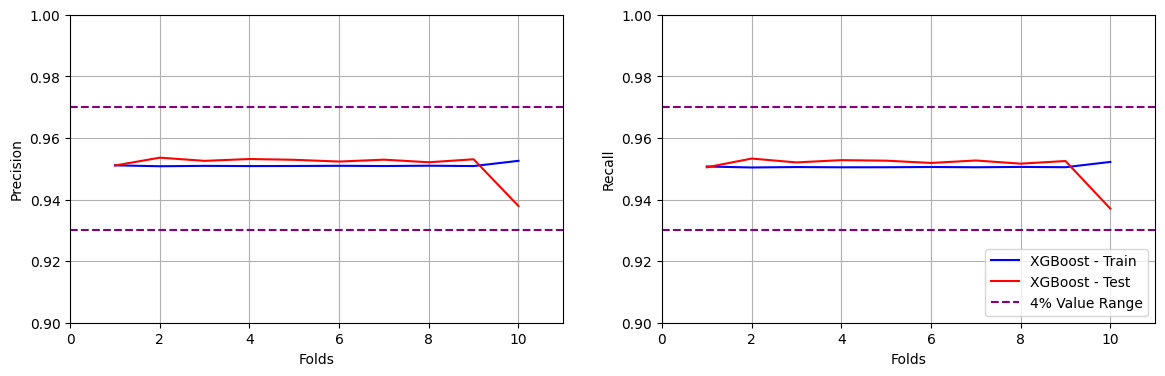}}
\caption{XGBoost Precision-Recall Stratified 10-Fold Cross Validation.}
\label{fig1}
\end{figure}

 The metrics with which we evaluated the performance of each algorithm are F1, Precision, Recall, and Roc-Auc. From the results in Tab.~\ref{table:1} and Tab.~\ref{table:2} we observe the high performance of ensemble and tree-based models, and more specifically of XGBoost, LightGBM, AutoGluon, XGBoost-RF, Random Forest, Extra Trees, and Decision Tree, while the baseline models, SVM and Logistic Regression fail. To highlight the impressive performance of the non-calibrated XGBoost, in terms of both stability and efficiency, we present the cross validation results for precision and recall in Fig.~\ref{fig1}.

\begin{table}[htbp]
  \centering
  \caption{Train/Test Precision and Recall Results ($\%$)}
  \label{table:1}
  \begin{tabular}{|c|c|c|c|c|}
    \hline
    \textbf{\textit{Model}} & \textbf{\textit{Train/Test-Precision}} & \textbf{\textit{Train/Test-Recall}} \\
    \hline
    LightGBM & 97.74/96.36 & 97.69/96.25 \\
    \hline
    AutoGluon & 92.34/93.53 & 91.32/92.30 \\
    \hline
    XGBoost & 98.67/96.40 & 98.64/96.29 \\
    \hline
    XGBoost-RF & 96.04/95.09 & 95.99/94.74 \\
    \hline
    Random Forest & 96.64/95.88 & 96.56/95.63 \\
    \hline
    HB-Gradient Boosting & 94.74/94.74 & 93.91/93.91 \\
    \hline
    Gradient Boosting & 71.82/75.17 & 71.03/73.47 \\
    \hline
    Extra Trees & 97.05/96.30 & 96.91/96.09 \\
    \hline
    Decision Tree & 96.92/95.28 & 96.85/95.03 \\
    \hline
    SVM & 7.08.5.57 & 0.32/3.41 \\
    \hline
    Logistic Regression & 8.16/6.40 & 6.47/6.82 \\
    \hline
  \end{tabular}
\end{table}
\begin{table}[htbp]
  \centering
  \caption{Train/Test F1 and Roc-Auc Results  ($\%$)}
  \label{table:2}
  \begin{tabular}{|c|c|c|c|c|}
    \hline
    \textbf{\textit{Model}} & \textbf{\textit{Train/Test-F1}} & \textbf{\textit{Train/Test-Roc-Auc}} \\
    \hline
    LightGBM & 97.69/96.25 & 98.81/98.08 \\
    \hline
    AutoGluon & 91.12/92.31 & 95.56/9606 \\
    \hline
    XGBoost & 98.65/96.33 & 99.30/98.11 \\
    \hline
    XGBoost-RF & 95.99/94.86 & 98.95/97.32 \\
    \hline
    Random Forest & 95.56/95.49 & 96.56/95.63 \\
    \hline
    HB-Gradient Boosting & 94.66/94.55 & 97.28/97.18 \\
    \hline
    Gradient Boosting & 69.14/72.78 & 85.18/86.64 \\
    \hline
    Extra Trees & 96.92/96.15 & 98.42/98.00 \\
    \hline
    Decision Tree & 96.85/95.11 & 98.37/97.46 \\
    \hline
    SVM & 1.29/1.34 & 50.45/50.50 \\
    \hline
    Logistic Regression & 5.05/4.84 & 52.15/52.16 \\
    \hline
  \end{tabular}
\end{table}

\section{Generation of Synthetic Data}
The generation of synthetic data has gained popularity now-days, especially with the hype of GenAI. It is very interesting how the generation of fake data can trick a machine learning based intrusion detection system, but even more interesting how we can use such generation techniques to construct synthetic datasets, that share statistical properties with the real ones. 

\subsection{Short Review of the Architecture of Generative Adversarial Networks and their Loss Functions}
A generative adversarial network architecture is consisted of two algorithms competing each other in a minmax game \cite{b18}. The first algorithm is called Generator, whose role is to generate synthetic data identical to the real ones and is commonly noted as $G$. The second one is called Discriminator, whose role is to distinguish between real and synthetic data and is commonly noted as $D$. For the most simple architectures $D$ and $G$ are neural networks. Let a training data-set of dimensions $n\times p$, where $n$ is the number of observations and $p$ is the number of features, while each record $\textbf{x}_{i}\in \mathcal{X}^p$, $i = 1, 2, ..., n$. The generator takes as input random noise, let $\textbf{z}\in R^k$, from a $k$-dimensional distribution and maps it to a $p$-dimensional space, thus generating a new fake data-point, let $x_{fake}^{(k)} = G(\textbf{z})\in \mathcal{X}^p$, $k = 1, 2, ..., m$, where $m$ is the number of newly generated data-points. Now let $D(\textbf{x}_{i})\in[0,1]$ be the probability of the $i$-th data-point to be real in this binary classification task. To be able to generate new synthetic data, the generator must learn how to create data identical to the real ones. This happens through the minmax game, which is a process of optimizing the parameters of each algorithm, i.e. the weights of each neural network. Thus, given the parameters $\boldsymbol{\theta}_D, \boldsymbol{\theta}_G$ of the discriminator and the generator respectively we can notate them as $D(\textbf{x}_{i};\boldsymbol{\theta}_D)$ and $G(\textbf{z};\boldsymbol{\theta}_G)$ respectively \cite{b18}. The stepping-stone for optimizing the procedure is the binary cross-entropy risk function (BCE), a special occasion of the categorical cross-entropy risk function (CCE), when we address a task with two probable outcomes. 
\par Given the discrete probability mass distributions $p(\cdot)$ and $q(\cdot)$, for a total of $n$ data-points and $C$ outcomes, the formulation of the risk functions of categorical cross-entropy and binary cross-entropy are formulated as seen in ``$(1)$'' and ``$(2)$''\cite{b35}, \cite{b36}:
\begin{equation}
R_{CCE}\Big(p(x), q(x)\Big) = \frac{1}{n}\sum_{i=1}^{n}\sum_{j=0}^{C-1}p(x_{ij})\log_2\Big(q(y_{ij}))\Big)
\end{equation}
\begin{equation}
\begin{split}
R_{BCE}\Big(p(x), q(x)\Big) =\frac{1}{n}\sum_{i=1}^{n}p(x_{i0})\log_2q(y_{i0}) + \\
+ \frac{1}{n}\sum_{i=1}^{n}p(x_{i1})\log_2q(y_{i1}) =
\frac{1}{n}\sum_{i=1}^{n}p(x_{i0})\log_2q(y_{i0}) + \\
+ \frac{1}{n}\sum_{i=1}^{n}\Big(1-p(x_{i0})\Big)\log_2\Big(1-q(y_{i0})\Big) 
\end{split}
\end{equation}
The last formulation of the binary cross-entropy is the most commonly used and can be explained since $p(y_0) + p(y_1) = 1$, where $S_Y = \{y_0, y_1\}$, where $p(\cdot)$ is a discrete probability mass distribution, with two outcomes, such that $S_Y = \{y_0, y_1\}$. From probability theory we have the following definition for the expected value of a transformation $g(X)$ of a discrete random variable $X$, with $f_X(x)\in[0,1]$ as the probability mass function and $S_X= \{x_1, x_2\}$\cite{b3}:
\begin{equation}
\begin{split}
    E_{f_X(x)}\Big\{g(X)\Big\} = \sum_{x\in S_X}g(x)f_X(x) =\\
    = g(x_1)f_X(x_1)+ g(x_2)f_X(x_2) = \\ 
    = g(x_1)f_X(x_1) + g(x_2)\Big(1-f_X(x_1)\Big)
    \end{split}
\end{equation}
Moving on from the theoretical background to the computational approach, given $\textbf{X}\sim p(\textbf{x})$, $\textbf{Z}\sim q(\textbf{z)}$, where $p(\textbf{x}), q(\textbf{z})$ are two probability distributions, such that $p(\textbf{x}) = 1.0$ if $\textbf{x}$ is real, $p(\textbf{x}) = 0.0$ if $\textbf{x}$ is fake, one formulation of the risk function used in the Vanilla-GAN, from keras and pytorch library can be seen in ``$(4)$'' as follows \cite{b36}: 
\begin{equation}
\begin{split}
R_{Vanilla-GAN}\Big(D(\textbf{x};\boldsymbol{\theta}_D), G(\textbf{z};\boldsymbol{\theta}_G)\Big) =  \\
= \frac{1}{n}\sum_{i=1}^{n}E_{p(\textbf{x})}\Big\{\ln\Big(D(\textbf{x}_{i};\boldsymbol{\theta}_D)\Big)\Big\} + \\
+ \frac{1}{n}\sum_{j=1}^{n}E_{q(\textbf{z})}\Big\{\ln\Big(1-D(G(\textbf{z}_{j};\boldsymbol{\theta}_G);\boldsymbol{\theta}_D)\Big)\Big\} = \\
= \frac{1}{n}\sum_{i=1}^{n}\Big\{p(\textbf{x}_{i})\ln\Big(D(\textbf{x}_{i};\boldsymbol{\theta}_D)\Big) + \\ + \Big(1 - p(\textbf{x}_{i})\Big)\ln\Big(D(\textbf{x}_{i};\boldsymbol{\theta}_D)\Big)\Big\} + \\
+ \frac{1}{n}\sum_{j=1}^{n}\Big\{q(\textbf{z}_{j})\ln\Big(1-D(G(\textbf{z}_{j};\boldsymbol{\theta}_G);\boldsymbol{\theta}_D)\Big) + \\
+ \Big(1-q(\textbf{z}_{j})\Big)\ln\Big(1-D(G(\textbf{z}_{j};\boldsymbol{\theta}_G);\boldsymbol{\theta}_D)\Big)\Big\} = \\
= \frac{1}{n}\sum_{i=1}^{n}\ln\Big(D(\textbf{x}_{i};\boldsymbol{\theta}_D)\Big) + \\ + \frac{1}{n}\sum_{j=1}^{n}\ln\Big(1-D(G(\textbf{z}_{j};\boldsymbol{\theta}_G);\boldsymbol{\theta}_D)\Big)
\end{split}
\end{equation}
The minmax procedure optimizes the following risk function \cite{b7}, \cite{b13}, \cite{b36} and more specifically it maximizes with respect to the discriminator $D$ and minimizes with respect to the generator $G$ as:
\begin{equation}
\min_{G}\max_{D} R_{Vanilla-GAN}\Big(D(\textbf{x};\boldsymbol{\theta}_D), G(\textbf{z};\boldsymbol{\theta}_G)\Big)
\end{equation}
For the conditional architecture, we add a condition in the data, the target variable, which is categorical, after it is One-Hot encoded. In this architecture, the $i$-th input is not $\textbf{x}_i$ as before, but has the form $(\textbf{x}_i, \textbf{y}_i)$, where $\textbf{y}_i$ can be the encoded label for the $i$-th observation $\textbf{x}_i$. In the basic architecture of CGAN, the minimax procedure has the following form, as seen in \cite{b36}:
\begin{equation}
\min_{G}\max_{D} R_{CGAN}\Big(D\big((\textbf{x}, \textbf{y});\boldsymbol{\theta}_D\big), G\big((\textbf{z},\textbf{y});\boldsymbol{\theta}_G\big)\Big)
\end{equation}
Moving to the $f$-GAN architectures, also known as variational divergence minimization \cite{b33}, the objective function is based on the idea of the convex conjugate function $f^*(t) = \sup_{u\in dom_f}\{ut-f(u)\}$, the Fenchel conjugate, of a convex, lower-semicontinuous function $f(u) = \sup_{t\in dom_{f^*}}\{tu - f^*(t)\}$. The minimax game occurs between the generator and the discriminator, or otherwise the variational function and are noted as $D(\textbf{x};\boldsymbol{\theta}_D)$ (or $T_{\boldsymbol{\theta}_D}(\textbf{x)} = g_f(V_{\boldsymbol{\theta}_D}(\textbf{x}))$) and $G(\textbf{z};\boldsymbol{\theta}_G)$ respectively. Given an activation function $g_f(\cdot)$, related to the conjugate $f^*(\cdot)$, for $V_{\boldsymbol{\theta}_D}(\textbf{x})$, the objective function, of the minmax process is formulated as seen is ``$(7)$'':
\begin{equation}
\begin{split}
F = \bigg\{ E_{p(\textbf{x})}\Big(D(\textbf{x};\boldsymbol{\theta}_D)\Big) -  E_{q(\textbf{x})}\Big( f^*\big(G(\textbf{z};\boldsymbol{\theta}_G)\big)\Big)\bigg\}
\end{split}
\end{equation}
The generator functions $f(u)$, along with their conjugate $f^*(t)$, for each $f$-divergence \cite{b33}, which we used can be seen in Tab.~\ref{table:3}:
\begin{table}[htbp]
  \centering
  \caption{$f$-GAN generator functions and their conjugates}
  \label{table:3}
  \begin{tabular}{|c|c|c|}
    \hline
    \textbf{\textit{Divergence}} & \textbf{\textit{$f(u)$}} & \textbf{\textit{$f^*(t)$}} \\
    \hline
    Kullback-Leibler & $ulog(u)$& $\exp\{t-1\}$\\
    \hline
    Squared-Hellinger & $\big(\sqrt{u}-1\big)^2$& $\frac{t}{1-t}$\\
    \hline
  \end{tabular}
\end{table}

\subsection{Training Adversarial Models}
For the generation of new synthetic data, we decided to select a wide range of different adversarial architectures. We chose Vanilla-GAN, Wasserstein-GAN (WGAN) \cite{b23}, Wasserstein-GAN with gradient penalty (WGAN-GP) \cite{b23}, $f$-GANs with Kullback-Leibler (KLD-GAN) and Squared-Hellinger divergence (H2D-GAN) \cite{b33}, the basic architecture of Conditional-GAN for tabular data (CTGAN-1) \cite{b12}, a Conditional-GAN equipped with Bayesian-Gaussian Mixture (CTGAN-2) \cite{b21}, a Variational Auto-Encoder (TVAE) \cite{b26}, an XGBoost-based Diffusion Forest (XGB-DF) \cite{b12}, the DistilGPT2 \cite{b12} and the PATE-CTGAN, a specialized privacy-preserving conditional generative adversarial network, which utilizes the private aggregation of teacher ensembles architecture (PATE) \cite{b15}. 
\par For the training of the adversarial networks we tested a few neural network architectures, we consider a total number of $20$ epochs for each of the above-mentioned algorithms. From the minimax procedure, for illustration purposes we can see below in Fig.~\ref{fig2} the discriminator's and generator's loss values for a few batches, for the Vanilla-GAN and WGAN, from which we observe the failure of Vanilla-GAN as the discriminator's loss converges to zero and the stable training WGAN, along with indications of convergence:
\begin{figure}[htbp]
\centerline{\includegraphics[width=0.95\columnwidth]{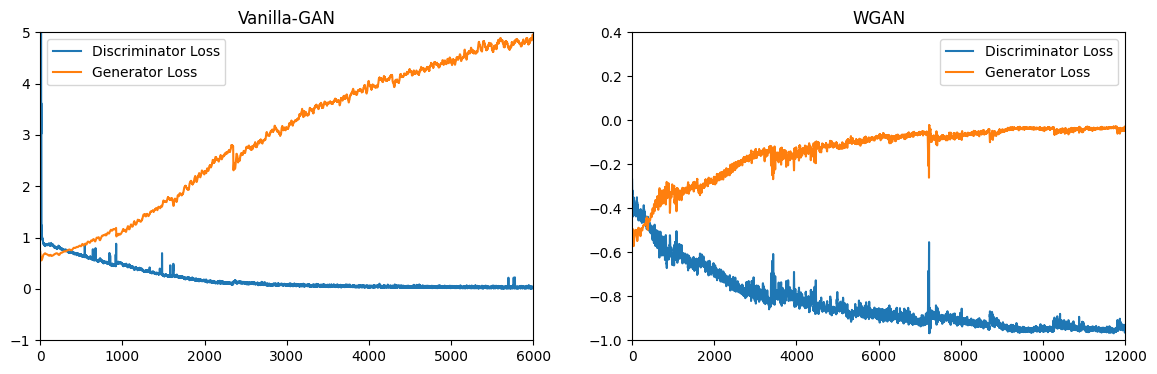}}
\caption{GAN and WGAN Loss Progress.}
\label{fig2}
\end{figure}
\par Furthermore we highlight the loss progress of TVAE, whose results are the highest, compared to the rest non-conditional architectures. Although it fails generating equal number of observations per network attack. It's performance could be further optimized with the use of a conditional architecture and a mixture approach:
\begin{figure}[htbp]
\centerline{\includegraphics[width=0.95\columnwidth]{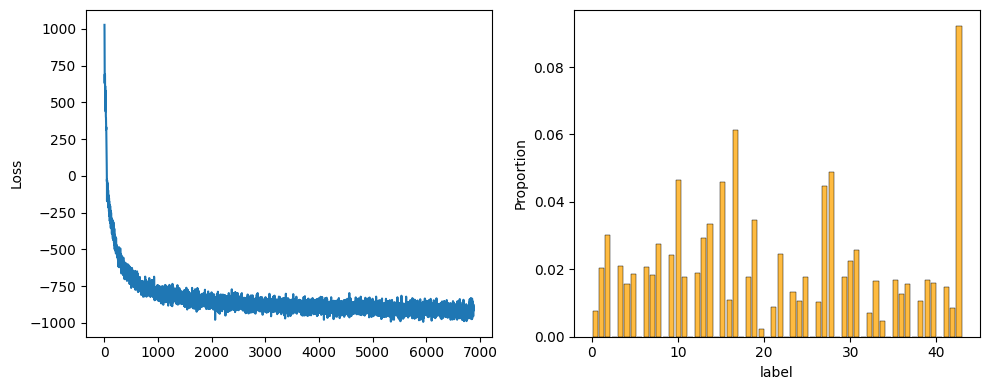}}
\caption{TVAE Results}
\label{fig3}
\end{figure}

\subsection{Evaluation}
For evaluating the quality of the new synthetically generated data we chose the Synthetic Data Vault framework (SDV) \cite{b26}. From the quality report, we investigated the statistical fidelity, measuring how well we have captured the marginal distribution of each feature using the Kolmogorov-Smirnov Complement and how well the correlations have been preserved using the Correlation Similarity, thus leading us to a final score which is the average of the two pre-mentioned scores. We also run the diagnostic report, in order to verify if the structural integrity and data validity have been maintained with the Table Structure score, which investigates if data-types per feature between the real and the synthetic are the same and the Boundary Adherence, which checks if the minimum-maximum boundaries per feature are respected. As before with the quality report we have a final score for the diagnostic report, which is the average of the Table Structure score and the Boundary Adherence score. For the computation of the above-mentioned metrics we used only the training set. We consider the $65\%$ and $95\%$ thresholds for the averaged Quality and Diagnostic report scores, to select a few algorithms to proceed to further evaluation. The averaged scores can be seen in Tab.~\ref{table:4}, where we highlight the performance of CTGAN-2 \cite{b21}, the diffusion forest and the LLM \cite{b12}.
\begin{table}[htbp]
  \centering
  \caption{SDV Overall Scores}
  \label{table:4}
  \begin{tabular}{|c|c|c|c|}
    \hline
    \textbf{\textit{Model}} & \textbf{\textit{Quality Report ($\%$)}} & \textbf{\textit{Diagnostic Report ($\%$)}} \\
    \hline
    Vanilla-GAN & 60.43 & 78.37 \\
    \hline
    WGAN & 57.07 & 81.24 \\
    \hline
    WGAN-GP & 63.55 & 78.72 \\
    \hline
    KLD-GAN & 63.55 & 78.72 \\
    \hline
    H2D-GAN & 49.85 & 80.22 \\
    \hline
    CTGAN-1 & 75.82 & 83.79 \\
    \hline
    CTGAN-2 & 98.91 & 100.00 \\
    \hline
    TVAE & 83.15 & 86.52 \\
    \hline
    XGB-DF & 99.93 & 100.00 \\
    \hline
    DistilGPT2 & 99.94 & 100.00 \\
    \hline
    PATE-CTGAN & 65.00 & 95.17 \\
    \hline
  \end{tabular}
\end{table}
\par After the selection of the models, we move forward with the Distinguishability Test, in order to further investigate if the synthetic data mimic the statistical properties of the real data. We consider two binary classifiers, a Random-Forest and the Logistic Regression as a baseline. From the baseline model results in Tab.~\ref{table:6} we observe that it decides randomly whether it's a real or a fake data-point since the Roc-Auc score is almost $50\%$ and the F1 score is $0\%$, for the first three models. For the differentially-privacy enhanced model we see that the baseline binary classifier, can't distinguish easily but tends to realize real from fake. As seen in Tab.~\ref{table:5} we see a good F1 score, fluctuating around $52\%$ for all models, but from the Roc-Auc we have indications that the Random Forest discriminator struggles, but slightly favors the real data, since the scores are close to $64\%$, which indicates that there is space for improvement. The only generative model, whose synthetic data can be easily distinguished by the random forest, is the PATE-CTGAN.
\begin{table}[htbp]
  \centering
  \caption{Random Forest - Distinguishability Test Results  ($\%$)}
  \label{table:5}
  \begin{tabular}{|c|c|c|}
    \hline
    \textbf{\textit{Model}} & \textbf{\textit{Test-F1}} & \textbf{\textit{Test-Roc-Auc}} \\
    \hline
    CTGAN-2 & 54.24 & 64.24 \\
    \hline
    XGB-DF & 53.39& 65.07\\
    \hline
    DistilGPT2 & 52.80 & 63.81 \\
    \hline
    PATE-CTGAN & 100.00 & 100.00 \\
    \hline
  \end{tabular}
\end{table}
\begin{table}[htbp]
  \centering
  \caption{Logistic Regression - Distinguishability Test Results  ($\%$)}
  \label{table:6}
  \begin{tabular}{|c|c|c|}
    \hline
    \textbf{\textit{Model}} & \textbf{\textit{Test-F1}} & \textbf{\textit{Test-Roc-Auc}} \\
    \hline
    CTGAN-2 & 0.00& 50.85\\
    \hline
    XGB-DF & 0.13& 52.80\\
    \hline
    DistilGPT2 & 0.20 & 53.70 \\
    \hline
    PATE-CTGAN & 14.65 & 63.18 \\
    \hline
  \end{tabular}
\end{table}
\par To further investigate the diversity and utility of the new generated data-sets and the scenario of mode collapse, we the perform Train-Real-Test-Synthetic (TRTS) and Train-Synthetic-Test-Real (TSTR) and compare the results with the Train-Real-Test-Real (TRTR) tests. For this comparison we selected the Random Forest classifier and as evaluation metrics the precision and recall score. In Tab.~\ref{table:7} we can find the precision and recall results for the synthetic data, which acted as a test set. From these scores we observe that they match the precision and recall scores for the original test data, apart from those of the PATE-CTGAN. From Tab.~\ref{table:8} we observe that the precision and recall scores of a newly trained Random-Forest, on the synthetic data, are fluctuating in the same levels of the scores seen in TRTR and TRTS. Though from PATE-CTGAN we have clear indications of poor data quality and lack of realism, for machine learning based tasks. All these results indicate high utility-quality data, with diversity and absence of mode collapse in CTGAN-2, XGB-DF and DistilGPT2.
\begin{table}[htbp]
  \centering
  \caption{TRTS Results ($\%$)}
  \label{table:7}
  \begin{tabular}{|c|c|c|}
    \hline
    \textbf{\textit{Model}} & \textbf{\textit{Precision}} & \textbf{\textit{Recall}} \\
    \hline
    CTGAN-2 & 96.59& 96.51\\
    \hline
    XGB-DF & 96.67& 96.59\\
    \hline
    DistilGPT2 & 96.67& 96.59\\
    \hline
    PATE-CTGAN & 00.54 & 0.02 \\
    \hline
  \end{tabular}
\end{table}
\begin{table}[htbp]
  \centering
  \caption{TSTR Results ($\%$)}
  \label{table:8}
  \begin{tabular}{|c|c|c|}
    \hline
    \textbf{\textit{Model}} & \textbf{\textit{Train/Test Precision}} & \textbf{\textit{Train/Test Recall}}\\
    \hline
    CTGAN-2 & 96.60/95.65& 96.51/95.38\\
    \hline
    XGB-DF & 96.76/95.56& 96.67/95.26\\
    \hline
    DistilGPT2 & 96.77/95.56& 96.67/95.26\\
    \hline
    PATE-CTGAN & 75.22/00.18 & 11.27/3.60 \\
    \hline
  \end{tabular}
\end{table}
\par We also investigated how well each generative model has captured the distribution of the data, by comparing them with a few divergence measures \cite{b18}, as seen below in Tab.~\ref{table:9}. Almost all models achieved, at a significant level, to generate sufficient samples from each attack category, thus capturing the underlying univariate distribution of the target variable.
\begin{table}[htbp]
  \centering
  \caption{Divergence-Based Evaluation Metrics}
  \label{table:9}
  \begin{tabular}{|c|c|c|c|}
    \hline
    \textbf{\textit{Divergence}} & \textbf{\textit{Jensen-Shannon}} & \textbf{\textit{Hellinger}} & \textbf{\textit{Wasserstein}} \\
    \hline
    CTGAN-2& 0.000967 & 0.031219 & 0.129161 \\
    \hline
    XGB-DF & 3.541914e-10 & 1.881997e-5 & 0.000106 \\
    \hline
    DistilGPT2 & 3.541914e-10 & 1.881997e-5 & 0.000106 \\
    \hline
    PATE-CTGAN & 0.004467 & 0.066443 & 0.729779  \\
    \hline
  \end{tabular}
\end{table}
\par For illustration purposes we show below in Fig.~\ref{fig4} and Fig.~\ref{fig5} the relative frequency per attack that was generated from CTGAN-2, the PATE-CTGAN and DistilGPT2, compared to the real proportions from the training set.
\begin{figure}[htbp]
\centerline{\includegraphics[width=0.95\columnwidth]{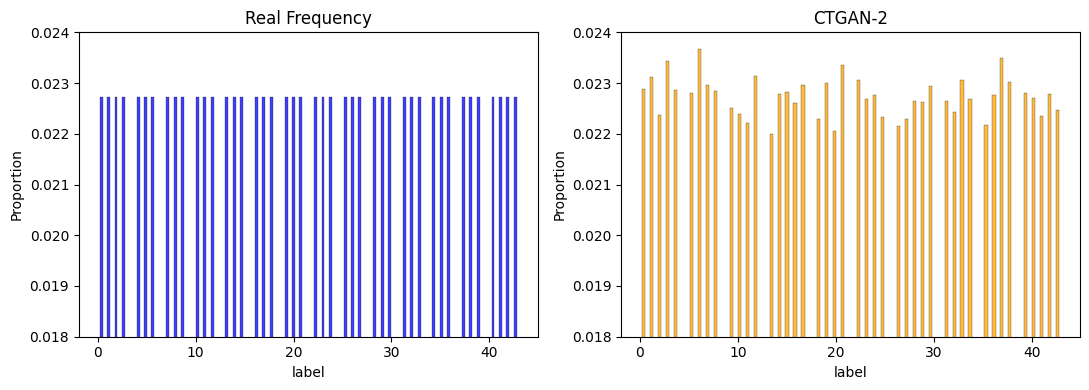}}
\caption{}
\label{fig4}
\end{figure}
\begin{figure}[htbp]
\centerline{\includegraphics[width=0.95\columnwidth]{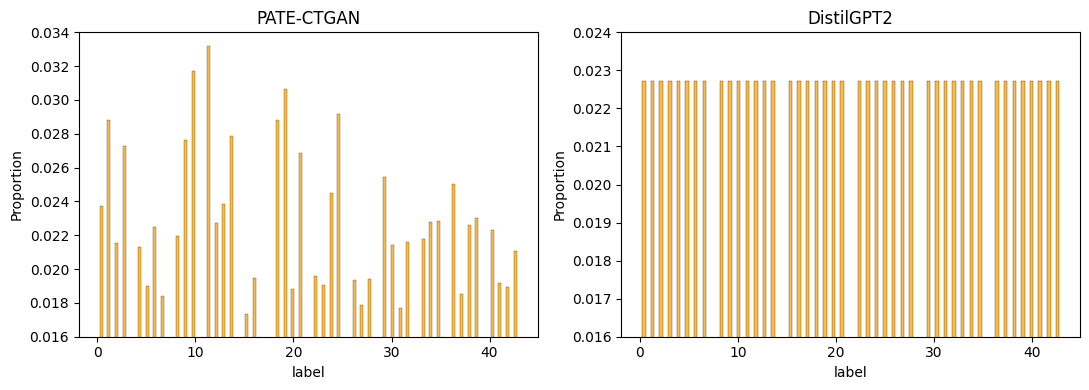}}
\caption{}
\label{fig5}
\end{figure}
\par We also performed statistical hypothesis tests in order to compare the multivariate means and the covariance matrices between the real and the synthetic data. We utilized the Regularized Hotelling's $T^2$ test \cite{b19} and the Frobenius Norm Covariance test \cite{b14}, \cite{b16}, for the multivariate means and covariance test respectively, each of which was computed for a total of $500$ permutations, to conclude to the $p-values$ seen in Tab.~\ref{table:10}. We consider a significance level of $a = 5\%$, for both statistical tests seen in ``$(8)$'' and ``$(9)$''. Regarding the equality of means, from all the below $p-values$, as seen in Tab.~\ref{table:10}, we have evidence to reject the alternative hypothesis $H_1$, at a significance level $a = 0.05$, since $p-value > a$, which leads us into the conclusion that the difference between the multivariate means of the real and synthetic data  are not statistically significant. From the $p-values$ of the Frobenius test, as seen in Tab.~\ref{table:11}, we have evidence to reject the alternative hypothesis, at a significance level $a = 0.05$, since $p-value > a$, in all tests, except the one considering the CTGAN-2 model, from which we have clear indications to reject the null hypothesis $H_0$, at a significance level $a = 0.05$, since $p-value < a$, meaning the difference between the covariance matrices of the real and synthetic data is statistically different at a significance level of $a = 0.05$.
\begin{equation}
    H_0:\boldsymbol{\mu}_1 = \boldsymbol{\mu}_2 \hhhh against \hhhh H_1:\boldsymbol{\mu}_1\neq \boldsymbol{\mu}_2
\end{equation}
\begin{equation}
    H_0:\boldsymbol{\Sigma_1} = \boldsymbol{\Sigma_2} \hhhh against \hhhh H_1:\boldsymbol{\Sigma_1}\neq \boldsymbol{\Sigma_2}
\end{equation}
\begin{table}[htbp]
  \centering
  \caption{Regularized Hotelling's $T^2$}
  \label{table:10}
  \begin{tabular}{|c|c|}
    \hline
    \textbf{\textit{Model}} & \textbf{\textit{p-values}} \\
    \hline
    CTGAN-2 & 0.99 \\
    \hline
    XGB-DF & 1.00 \\
    \hline
    DistilGPT2 & 1.00 \\
    \hline
    PATE-CTGAN & 0.99 \\
    \hline
  \end{tabular}
\end{table}
\begin{table}[htbp]
  \centering
  \caption{Frobenius Norm Covariance Test Results}
  \label{table:11}
  \begin{tabular}{|c|c|}
    \hline
    \textbf{\textit{Model}} & \textbf{\textit{p-values}} \\
    \hline
    CTGAN-2 & 0.0019 \\
    \hline
    XGB-DF & 0.9381\\
    \hline
    DistilGPT2 & 0.9261 \\
    \hline
    PATE-CTGAN & 0.14 \\
    \hline
  \end{tabular}
\end{table}
\par We applied a two-sided non-parametric statistical test, Maximum Mean Discrepancy (MMD) \cite{b17}, \cite{b4} to measure the distance between the joint distribution of the features of the real and synthetic datasets, with $500$ permutations in randomized samples of the original real and synthetic datasets, which computes the mean embeddings in a Reproducing Kernel Hilbert Space (RKHS). We consider a significance level of $a = 5\%$ and their $p-values$ can be seen in Tab.~\ref{table:12}, for each generative model. From the table, we observe that we have no evidence to reject the null hypothesis ($H_0:P = Q$) at a significance level of $a = 5\%$, apart from the PATE-CTGAN model, which eventually failed to capture the multivariate distribution, mostly because of the differential privacy constraints. Thus, we can't assume that the difference between their distributions is statistically different for the CTGAN-2, the diffusion forest and the LLM.
\begin{table}[htbp]
  \centering
  \caption{MMD Test Results}
  \label{table:12}
  \begin{tabular}{|c|c|c|}
    \hline
    \textbf{\textit{Model}} & \textbf{\textit{p-values}} \\
    \hline
    CTGAN-2 & 0.0829  \\
    \hline
    XGB-DF & 1.0000 \\
    \hline
    DistilGPT2 & 1.0000  \\
    \hline
    PATE-CTGAN & 0.0000 \\
    \hline
  \end{tabular}
\end{table}
\par For the scenario of data leakage we use a privacy evaluation score for the generated data by each of the selected models \cite{b11}, \cite{b28}, to check if they are simply copying the existing records instead of synthesizing new ones, but also weight the trade-off between privacy and utility-quality. We utilized the Nearest Neighbor Distance Record score (NNDR). From Tab.~\ref{table:13} we observe the NNDR scores between the train and test sets compared to the synthetic datasets, to measure overfitting, while considering a threshold of $4\%$ for the score difference between the train and the test sets. From the below scores, we have good NNDR scores, showing standard to high levels of privacy, especially for the PATE-CTGAN, but we have indications of overfitting for the other models, which don't have a privacy-preserving architecture. 
\begin{table}[htbp]
  \centering
  \caption{NNDR Results}
  \label{table:13}
  \begin{tabular}{|c|c|c|}
    \hline
    \textbf{\textit{Model}} & \textbf{\textit{Train-NNDR}} & \textbf{\textit{Test-NNDR}} \\
    \hline
    CTGAN-2 & 0.6136&  0.7171\\
    \hline
    XGB-DF & 0.6122& 0.7194\\
    \hline
    DistilGPT2 & 0.6140&  0.7193\\
    \hline
    PATE-CTGAN & 0.9634 & 0.9731 \\
    \hline
  \end{tabular}
\end{table}

\section{Conclusions}
In this paper we clarified how efficient, in terms of speed, accuracy and stability, the tree and ensemble models are in classifying network attacks compared to other baseline models, when the network data are processed in tabular format. The conditional architectures, when combined with mixture models have outstanding performance in capturing the probability distribution function of the conditional variable, contrast to basic conditional architectures. We also highlight the diffusion and LLM models for their performance in the distinguishability tests and for capturing the underlying distributions and the quality-privacy tradeoff. The use of robust and non-parametric statistical tests for analyzing the multivariate nature of the synthetic data, generated either from non-differential or differential architectures give us a solid perspective of the joint distribution of the features. Future work includes cross-dataset validation with other well known IDS datasets, using adversarial attacks for this multi-modal dataset, using conditional, privacy-preserving architectures for generative models, with Bayesian mixtures, for the unbalanced version of the UMNIDS dataset. Further evaluation is necessary with different differential-privacy architectures and their effect in privacy scores in combination with fidelity.

\section{Acknowledgments}
This work has been partially funded by the ‘advaNced cybErsecurity awaReness ecOsystem for SMEs’ (Nero) project, which has received funding from the European Union’s Digital Europe Programme (DEP) under grant agreement No. 101127411. Prof. Christos Douligeris would like to thank Prof. E. Iakovou and the Energy Institute of the Texas A$\&$M University for their hospitality.

\end{document}